\documentclass[12pt]{article}

\usepackage[tmargin=2cm,lmargin=2.5cm,bmargin=3cm,hmarginratio=1:1,headheight=65pt,headsep=1cm]{geometry} 
\usepackage[T1]{fontenc}
\usepackage[french,english]{babel}
\usepackage{color,graphicx}
\usepackage{amsmath,amssymb,amsfonts}
\usepackage{amsthm}
\usepackage{mathtools}
\usepackage{slashed}
\usepackage{sidecap} 
\usepackage{array}
\usepackage[indentafter]{titlesec}
\usepackage[utf8]{inputenc}
\usepackage{setspace}
\usepackage{perpage} 
\usepackage{fancyhdr} 
\usepackage{emptypage}
\usepackage{pifont} 
\usepackage{cite} 
\usepackage{epsfig,latexsym,amsfonts,amsmath,amsthm,amssymb,amsbsy,multirow,slashed,wasysym,textcomp,subfigure,hyperref,wrapfig,datetime,comment,cancel}

\usepackage{hyperref} 

\newcommand{\p}{\partial}
\newcommand{\rd}{\mathcal}

\newcommand{\be}{\begin{eqnarray}}
\newcommand{\en}{\end{eqnarray}}
\newcommand{\badat}{\begin{alignedat}}
\newcommand{\eadat}{\end{alignedat}}
\newcommand{\bitm}{\begin{itemize}}
\newcommand{\eitm}{\end{itemize}}
\newcommand{\bmat}{\begin{pmatrix}}
\newcommand{\emat}{\end{pmatrix}}
\newcommand{\ba}{\begin{align}}
\newcommand{\bas}{\begin{align*}}
\newcommand{\ab}{\end{align}}
\newcommand{\bse}{\begin{subequations}}
\newcommand{\ese}{\end{subequations}}

\newcommand{\virg}{\hspace{1 mm}, \hspace{8 mm}}

\def\ndelta{\delta\hspace{-0.50em}\slash\hspace{-0.05em} }

\numberwithin{equation}{section} 

\begin{document}

\begin{titlepage}
  \thispagestyle{empty}

  \begin{flushright}
  CPHT-RR013.032019
  \end{flushright}

\vskip3cm

  \begin{center}  
{\LARGE\textbf{Carrollian Physics at the Black Hole Horizon}}

\vskip1cm

   \centerline{Laura Donnay$^{a,b}$, Charles Marteau$^{c}$}

\vskip1cm

{$^a$ Center for the Fundamental Laws of Nature, Harvard University}\\
{{\it 17 Oxford Street, Cambridge, MA 02138, USA.}}

{$^b$ Black Hole Initiative, Harvard University}\\
{{\it 20 Garden Street, Cambridge, MA 02138, USA.}}

{$^c$ CPHT,
	Ecole Polytechnique, CNRS UMR 7644,
	Universit\'e Paris-Saclay}\\
{{\it 91128, Palaiseau, France.}}

\end{center}

\vskip1cm

\begin{abstract}
We show that the geometry of a black hole horizon can be described as a Carrollian geometry emerging from an ultra-relativistic limit where the near-horizon radial coordinate plays the role of a virtual velocity of light tending to zero. 
We prove that the laws governing the dynamics of a black hole horizon, the null Raychaudhuri and Damour equations, are Carrollian conservation laws obtained by taking the ultra-relativistic limit of the conservation of an energy--momentum tensor; we also discuss their physical interpretation. We show that the vector fields preserving the Carrollian geometry of the horizon, dubbed Carrollian Killing vectors, include BMS-like supertranslations and superrotations and that they have non-trivial associated conserved charges on the horizon. In particular, we build a generalization of the angular momentum to the case of non-stationary black holes. Finally, we discuss the relation of these conserved quantities to the infinite tower of charges of the covariant phase space formalism.
\end{abstract}

\end{titlepage}

\tableofcontents

\section{Introduction}\label{I}
In the membrane paradigm formalism \cite{MacdonaldD.1982Beaa,Thorne:1986iy,Price:1986yy}, the black hole event horizon is seen as a two-dimensional membrane that lives and evolves in three-dimensional spacetime. This viewpoint was originally motivated by Damour's seminal observation that a generic black hole horizon is similar to a fluid bubble with finite values of electrical conductivity, shear and bulk viscosity \cite{DamourCurrents,Damour,DamourProc}. It was moreover shown that the equations governing the evolution of the horizon take the familiar form of an Ohm's law, Joule heating law, and Navier-Stokes equation. The membrane paradigm developed by Thorne and Macdonald for the electromagnetic aspects, and by Price and Thorne for gravitational and mechanical aspects, combines Damour's results with the $3+1$ formulation of general relativity, where one trades the true horizon for a 2+1-dimensional timelike surface located slightly outside it, called ``stretched horizon'' or ``membrane''. The laws of evolution of the stretched horizon then become boundary conditions on the physics of the external universe, hence making the membrane picture a convenient tool for astrophysical purposes. In order to derive the evolution equations of the membrane, a crucial step in \cite{Price:1986yy} was to renormalize all physical quantities (energy density, pressure, etc) on the membrane, as they turned out to be divergently large as one approaches the real horizon. We will show that a better approach to this issue is to interpret the near-horizon limit as an ultra-relativistic limit for the stretched horizon, where the radial coordinate plays the role of a virtual speed of light. This ultra-relativistic limit results in the emergence of Carrollian physics at the horizon.

The Carroll group was originally introduced in \cite{Leblond} as an ultra-relativistic limit of the Poincaré group where the speed of light is tending to zero (as opposed to the more familiar non-relativistic limit leading to the Galilean group). Recently, there has been a renewed interest in Carrollian physics due to its relation to asymptotically flat gravity. The symmetry group of asymptotically flat spacetimes is the Bondi-Metzner-Sachs (BMS) group; it is an infinite-dimensional extension of the Poincaré group, and its connection with soft theorems has shead a new light on the infrared structure of gravitational theories \cite{Strominger:2017zoo}. Interestingly, the BMS group was also shown to be isomorphic to a conformal extension of the Carroll group in \cite{Duval:2014uva}, while the dynamics of asymptotically flat spacetimes has been rephrased in terms of ultra-relativistic conservations laws on null infinity \cite{Ciambelli:2018ojf}. This leads to think that theories holographically dual to asymptotically flat gravity should be ultra-relativistic and enjoy a Carrollian symmetry \cite{Bagchi:2019xfx}. Actually, it is now understood that \textit{any} null hypersurface is endowed with a Carrollian geometry\footnote{Carrollian geometry is the degenerate geometry that one obtains when taking the ultra-relativistic limit of a Lorentzian metric. It is composed of a non-degnerate metric on spatial sections and a transverse vector field that corresponds to the time direction.} \cite{Duval:2014uoa,Duval:2014lpa,Hartong:2015xda,Blau:2015nee,Ciambelli:2018xat,Ciambelli:2018ojf} and that the associated constraint equations are ultra-relativistic conservation laws \cite{Charlestoappear}. The aim of this paper is to give a complete analysis of this statement at the level of another physically interesting null hypersurface, the horizon of a black hole. The Carrollian symmetry emerging at the horizon was also used in  \cite{Penna:2018gfx} to explain the vanishing of Love numbers for the Schwarzschild black hole.

The recent focus on the symmetries of near-horizon geometries has been motivated by the fact that they exhibit, in some instances, a BMS-like algebra composed of supertranslations and superrotations \cite{Donnay:2015abr,Donnay:2016ejv,Donnay:2018ckb,Hawking:2016msc,Hawking:2016sgy,Haco:2018ske,Haco:2019ggi,Afshar:2016kjj,Grumiller:2018scv,Penna:2015gza,Penna:2017bdn,Chu:2018tzu,Kirklin:2018wvq,Chandrasekaran:2018aop}. Moreover, one can associate non-trivial charges to these large diffeomorphisms: they generate the so-called \textit{soft hair} on black holes \cite{Hawking:2016msc,Hawking:2016sgy,Haco:2018ske,Haco:2019ggi}, which were pointed out to have implications for the information paradox. We will show that this rich symmetry structure is in fact naturally encoded in the  Carrollian geometry of the horizon. To do so, we will interpret the near-horizon limit as an ultra-relativistic limit, where the radial coordinate $\rho$ plays the role of a virtual speed of light for constant $\rho$ hypersurfaces. This will allow to define proper, rather than ad hoc, finite quantities on the horizon. Moreover, we will prove that the laws governing the dynamics of the black hole horizon are Carrollian conservation laws. These are the ultra-relativistic equivalent of the conservation of an energy--momentum tensor. Through the near-horizon analysis, we will derive the isometries of the induced Carrollian geometry on the horizon and show that they include supertranslations and superrotations. We will also construct associated conserved charges; in particular, the one associated with superrotations will provide a generalization of the angular momentum for very generic non-stationary black holes. Finally, the relation of these conserved quantities to the charges of the covariant phase space formalism will also be discussed.

The paper is organized as follows: in Sec. \ref{II}, we introduce a suitable coordinate system for the study of near-horizon geometries. We define the intrinsic and extrinsic objects of the horizon and write the constraint equations governing the dynamics, \emph{i.e.} the null Raychaudhuri and Damour equations. We then review the set of vector fields preserving the near-horizon metric and the derivation of their associated surface charges defined in the covariant phase space formalism. In Sec. \ref{III}, we present the Carrollian geometry associated with the black hole horizon. By identifying the radial coordinate $\rho$ as the square of a virtual speed of light for constant $\rho$ hypersurfaces, we interpret the near-horizon limit ($\rho\rightarrow 0$) as an ultra-relativistic limit and compute the horizon Carrollian geometric fields. We then define the energy--momentum tensor associated with a constant $\rho$ hypersurface in terms of its extrinsic curvature. The analysis of its scaling w.r.t. the radial coordinate allows us to define the Carrollian momenta which are the ultra-relativistic equivalent of the energy--momentum tensor. We give a physical interpretation of those quantities in terms of energy density, pressure, heat current and dissipative tensor. Ultra-relativistic conservation laws are written in terms of the Carrollian momenta and are shown to match perfectly the null Raychaudhuri and Damour equations. Finally, we consider the Killing fields which preserve the Carrollian geometry induced on the horizon and construct associated conserved charges.  The latter provides a generalization of the angular momentum for non-stationary black holes. We extend this analysis to conformal Killing vectors of the Carrollian geometry and show that the charges are now conserved provided a conformal state equation involving the energy density and the pressure is satisfied. We also write an interesting relation between these conserved charges and the one obtained in the covariant phase space formalism. We conclude in Sec. \ref{IV} with a discussion of open questions.

\section{Near-horizon geometry and dynamics}\label{II}
In this section, we describe the near-horizon geometry of a black hole and its dynamics. To do so, we introduce a coordinate system adapted to the study of the spacetime geometry near a null hypersurface. This will allow us to define the intrinsic and extrinsic geometry of the horizon. The projection of Einstein equations on the horizon gives rise to two constraint equations on the extrinsic geometry, the \emph{null Raychaudhuri} equation and the \emph{Damour} equation. These are the constraints that we ultimately want to interpret as ultra-relativistic conservation laws. Finally, we turn to the asymptotic symmetries preserving the form of the near-horizon geometry we have introduced, and present the associated charges computed through the covariant phase space formalism. They have the particularity of being generically non-integrable.

\subsection{Intrinsic and extrinsic geometry of the horizon}\label{section2.1}

We consider a $D$-dimensional spacetime whose coordinates are $x^a=(x^{\alpha},x^A)$, with $x^\alpha=(v,\rho)$ where $v$ is the advanced time and $\rho$ the radial coordinate. The surfaces of constant $v$ and $\rho$ are $(D-2)$-dimensional spheres $S_{v,\rho}$ and parametrized by $x^A$ ($A=3,\cdots,D$), the set of all these angular coordinates will be denoted $\mathbf{x}$. Throughout the paper, when we refer to spatial objects, it will be with respect to the angular coordinates. The constant $v$ surfaces are null, and constant $\rho$ are timelike. Finally, we assume the existence of a horizon $\mathcal{H}$ sitting at $\rho=0$.

\begin{figure}[ht!]
\begin{center}
 \includegraphics[width=.5\textwidth]{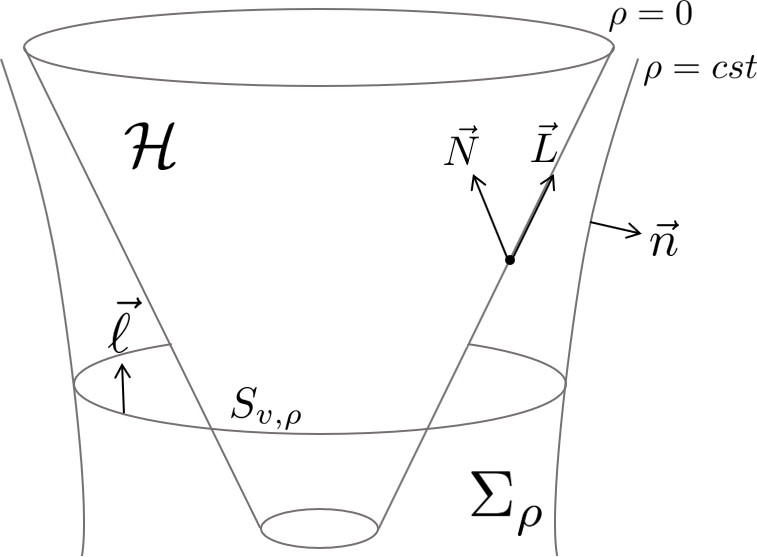} 
 \end{center}
 \caption{The horizon is a null hypersurface situated at $\rho=0$ and $\Sigma_{\rho}$ is a timelike constant $\rho$ hypersurface near the horizon. We define also four vectors that are useful for our analysis, the null vector $\vec{L}$ is the normal to the horizon while $\vec{N}$ is transverse but also null. The spacelike vector $\vec{n}$ is the normal to $\Sigma_{\rho}$ and the timelike vector $\vec{\ell}$ is the normal to a constant $v$ section of $\Sigma_{\rho}$.} 
 \label{ourfigure}
\end{figure}

It is alway possible to find a coordinates system, usually called \emph{null Gaussian coordinates}, such that the near-horizon geometry is given by  \cite{Moncrief:1983xua}\footnote{See also \cite{Chrusciel}, p. 48 for a review.}
\begin{equation}
ds^2=-2\kappa \rho dv^2+2d\rho dv + 2\theta_A \rho dv dx^A+(\Omega_{AB}+\lambda_{AB} \rho)dx^A dx^B+\mathcal{O}(\rho^2),
\label{metric}
\end{equation}
where $\kappa$, $\Omega_{AB}$, $\lambda_{AB}$, $\theta_A$ in principle depend on the coordinates $\mathbf{x}$ and $v$. The spatial metric $\Omega_{AB}$ will be used to raise and lower spatial indexes. We will sometime refer to the $D$-dimensional spacetime as the \emph{bulk}.

There are now two types of geometrical objects we can define on $\mathcal{H}$: the first ones are intrinsic and the others extrinsic. In a Hamiltonian perspective, they are canonical conjugate of each other. Moreover, the canonical momenta satisfy constraint equations that are imposed by the gravitational dynamics \cite{Hopfmuller:2016scf, Hopfmuller:2018fni}. The induced geometry on $\rd H$ is degenerate and reads
\begin{equation}\label{metricind}
 ds^2_{\rd H}=0 \cdot dv^2+0 \cdot dvdx^A+\Omega_{AB}dx^Adx^B,
\end{equation}  
the intrinsic geometry being then entirely specified by the spatial metric in this gauge. We now perform a decomposition of the bulk metric adapted to the study of null hypersurfaces:
\begin{equation}
g_{ab }=q_{ab }+L_{a }N_{b}+N_{b}L_{a },
\end{equation}
where
\begin{equation}
\vec{L}=L^a \p_a=\partial_v-\rho \theta^A \partial_A+\kappa \rho\partial_\rho\quad \text{and}\quad N=N_a dx^a=dv,
\end{equation}
are respectively a null vector and a null form. They satisfy $N(\vec{L})=1$ and will allow us to define all the extrinsic curvature elements of $\mathcal{H}$. The vector $\vec{L}$ coincides with the normal to the horizon on $\mathcal{H}$, and has the particularity of being also tangent to the horizon. On the other hand the vector $\vec{N}\equiv g^{-1}(N)$ is transverse to the horizon and together with $\vec{L}$ they define $q_{ab}$, the projector perpendicular to $\vec{L}$ and $\vec{N}$. In his work \cite{Damour,DamourProc}, T. Damour maps the black hole dynamics to the hydrodynamics of a fluid living on the horizon, and the vector $\vec{L}$ defines the fluid's velocity through $\vec{L}_{\mathcal{H}}=\partial_v+v^A\partial_A$. We have $v^A=0$, as we have chosen comoving coordinates, \emph{i.e.}, in Damour's interpretation the fluid would be at rest but on a dynamical surface\footnote{As pointed out in \cite{Price:1986yy}, one can always set $v^A=0$, namely the spatial coordinates $x^A$ can always be taken to be comoving, except at caustics.}.

The extrinsic geometry of the horizon is captured by a triple $(\Sigma^{AB}, \omega_{A},\tilde{\kappa})$ where $\Sigma^{AB}$ is the deformation tensor (or second fundamental form), $\omega_A$ is the twist field (Hajicek one-form) and $\tilde{\kappa}$ the surface gravity, defined as follows:
\begin{equation}
\begin{split}
\Sigma_{AB}=\frac{1}{2}q_A^a q_B^b\mathcal{L}_{\vec{L}}q_{ab},\quad
\omega_A=q_A^a(N_b \mathcal{D}_a  L^b)\quad\text{and}\quad
L^b\mathcal{D}_bL^a =\tilde{\kappa} L^a,
\end{split}
\end{equation}
where $\rd L$ denotes the Lie derivative, and $\mathcal{D}_a$ is the Levi-Civita associated with $g_{ab}$. Using the bulk metric \eqref{metric}, these quantities become on $\mathcal{H}$
\begin{equation}
\label{extrinsic}
\begin{split}
\Sigma_{AB}=\frac{1}{2}\partial_v\Omega_{AB},\quad
\omega_A=-\frac{1}{2}\theta_A\quad\text{and}\quad
\tilde{\kappa}=\kappa.
\end{split}
\end{equation}
We see that $\kappa$ really plays the role of the surface gravity and that $\theta_A$ is proportional to the twist. The deformation tensor gives rise to two new extrinsic objects:  its trace and its traceless part, which are respectively the horizon expansion and the shear tensor:
\begin{equation}
\begin{split}
&\Theta=\Omega^{AB}\Sigma_{AB}=\partial_v\ln\sqrt{\Omega},\\
&\sigma_{AB}=\frac{1}{2}\partial_v\Omega_{AB}-\frac{\Theta}{D-2}\Omega_{AB},
\end{split}
\label{shear}
\end{equation}
where $\sqrt{\Omega}$ is the volume form of the spatial metric. The scalar expansion $\Theta$ measures the rate of variation of the surface element of the spatial section of $\rd H$.\footnote{By definition, a non-expanding horizon has $\Theta=0$.}  It is possible to show, under the assumption that matter fields satisfy the null energy condition and that the null Raychaudhuri equation (see next section) is satisfied, that $\Theta$ is positive everywhere on $\rd H$, which implies that the surface area of the horizon can only increase with time (see e.g. \cite{Wald:106274}).

\subsection{Raychaudhuri and Damour equations}
Those quantities being defined, we can deduce from Einstein equations two conservation laws (or constraint equations) that belong to $\mathcal{H}$: the null Raychaudhuri equation \cite{Raychaudhuri:1953yv} and Damour equation \cite{Damour,DamourProc}, which are respectively
\begin{equation}
L^aL^bR_{ab}=0\quad \text{and}\quad q_A^a L^b R_{ab}=0;
\end{equation} 
they are thus given by projections of vacuum Einstein equations on the horizon. The first one is scalar and the second one is a vector equation w.r.t. the spatial section of $\mathcal{H}$. Using the near-horizon geometry \eqref{metric}, the null Raychaudhuri equation becomes
\begin{equation}\label{Ray}
\partial_v\Theta-\kappa\Theta+\frac{\Theta^2}{D-2}+\sigma_{AB}\sigma^{AB}=0,
\end{equation}
where $\sigma^{AB}=\Omega^{AC}\Omega^{BD}\sigma_{CD}$. This equation describes how the expansion evolves along the null geodesic congruence $\vec{L}$ and is a key ingredient in the proofs of singularity theorems. Damour equation\footnote{Raychaudhuri and Damour equations are called respectively the ``focusing equation''  and the ``Hajicek equation''  in Price-Thorne \cite{Price:1986yy}. The tidal-force equation expresses components of the Weyl tensor in terms of the evolution of the shear, and will not play a role here.} becomes
\begin{equation}\label{Damour}
\left(\partial_v+\Theta\right)\theta_A+2\nabla_A\left(\kappa+\frac{D-3}{D-2}\Theta\right)-2\nabla_B\sigma^B_A=0,
\end{equation}
where $\nabla_A$ is the Levi-Civita connection associated with $\Omega_{AB}$. 
Damour has interpreted this last equation as a $(D-2)$-dimensional Navier-Stokes equation for a viscous fluid; notice that the fluid velocity is not appearing here because we have chosen a comoving coordinate system as explained earlier. 

It is useful to know what these equations become when considering the conformal gauge, \emph{i.e.} when the spatial metric can be written as a conformal factor times a purely spatial metric:
\begin{equation}
\Omega_{AB}=\gamma(v,\mathbf{x})\bar{\Omega}_{AB}(\mathbf{x}).
\end{equation}
One can check that this is equivalent to asking the shear to be zero. If we make this choice, $\bar{\Omega}_{AB}$ disappears and the two conservation equations read
\begin{equation}
\begin{split}\label{DD}
&\partial_v^2\gamma-\frac{1}{2}\gamma^{-1}(\partial_v\gamma)^2-\kappa\partial_v\gamma=0,\\
&\partial_v\theta_A+2\partial_A\kappa+(D-3)\gamma^{-1}\partial_A\partial_v\gamma-(D-3)\gamma^{-2}\partial_A\gamma\partial_v\gamma+\frac{(D-2)}{2}\gamma^{-1}\theta_A\partial_v\gamma=0.
\end{split}
\end{equation}
In particular, one can verify that these equations reproduce the field equations studied in \cite{Donnay:2016ejv} in the $D=3$ and $D=4$ cases.

\subsection{Bulk symmetries and associated charges}
\label{sec:bulkcharges}

We now turn our attention to the bulk symmetries of the near-horizon gauge. The vector fields $\chi=\chi^a \p_a$ that preserve the shape of the metric \eqref{metric} were shown in \cite{Donnay:2016ejv} to involve of a smooth arbitrary function $f(v,\bold x)$, which depends on the advanced time and the sphere coordinates, and a vector field of the sphere $Y^A(\bold x)$; they are given by 
\begin{equation}
\begin{split}
&\chi^v=f(v,\bold x),\\
&\chi^{\rho}=-\partial_vf\rho+\frac{1}{2}\theta^A\partial_Af\rho^2+\mathcal{O}(\rho^3),\\
&\chi^A=Y^A(\bold x)+\Omega^{AC}\partial_Cf\rho+\frac{1}{2}\lambda^{AC}\partial_Cf\rho^2+\mathcal{O}(\rho^3),
\end{split}
\label{Killing}
\end{equation} 
and in any dimension $D$. We will call them asymptotic Killing vectors even though the gauge introduced does not involve a notion of infinity. We notice an important feature, which is that these vector fields projected on the horizon become
\begin{equation}
\chi=f(v,\bold x)\partial_v+Y^A(\bold x)\partial_A\quad\text{projected on $\mathcal{H}$},
\end{equation}
and as $f$ and $Y^A$ are totally generic for the moment, this is exactly the infinitesimal version of a particular type of diffeomorphisms on the horizon that we will define in Sec. \ref{III}: the Carrollian diffeomorphisms. Following \cite{Donnay:2015abr,Donnay:2016ejv}, we will call $f$ a supertranslation and $Y^A$ a superrotation. They act on the horizon fields in the following way:
\begin{equation}
\badat{3}\label{deltachi}
&\delta_{\chi} \kappa =Y^{A} \p_{A} \kappa + \p_v (\kappa f) + \p_v^2 f,  \\ 
&\delta_{\chi}  \Omega_{AB} = f\partial_v\Omega_{AB} + {\mathcal L}_{Y}\Omega_{AB},\\
&\delta_{\chi} \theta_A =\rd L_Y \theta_{A}+ f \partial_v\theta_A -2\kappa \partial_A f -2\partial_v\partial_A f + \partial_v\Omega_{AB} \partial^B f ,\\
&\delta_{\chi}  \lambda_{AB}=f\partial_v \lambda_{AB} -\lambda_{AB}\partial_v f +\rd L_Y \lambda_{AB} +\theta_A\partial_B f+\theta_B\partial_A f -2\nabla_A\nabla_B f.
\eadat
\end{equation}

To each of these vector fields preserving the near-horizon metric, one can associate a surface charge through the covariant phase space formalism \cite{Barnich:2001jy}.\footnote{See also \cite{Compere:2018aar} for a pedagogical introduction to this formalism.} More precisely, the quantity which is constructed at first is not a charge, but rather the field-variation of a charge (namely a one-form in the configuration space). For an on shell metric $g$ and variation $h\equiv\delta g$, it is given by:
\begin{equation} \label{deltaQ}
\ndelta Q_\chi[g,h]=\oint_{S_{v,\rho}}\mathbf{k}_{\chi}[g,h],
\end{equation}
where $\chi$ is an asymptotic Killing vector and $\mathbf{k}_{\chi}[g,h]$ is a one-form w.r.t. the field configuration space but a $(D-2)$-form w.r.t. the spacetime. It is defined as follows:\footnote{There is actually an ambiguity in this definition (see \cite{Compere:2018aar}), as one can add to the definition of $\mathbf{k}_{\chi}[g,h]$ the term $\alpha\frac{\sqrt{-g}}{16\pi G}(d^{D-2}x)_{ab}\left(h^{cb}\nabla^a\chi_c+h^{cb}\nabla_c\chi^a\right)$, where $\alpha$ is any constant. But one can show that for the metric and vector fields at hand, this term vanishes when evaluated on the horizon.}
\begin{equation}
\mathbf{k}_{\chi}[g,h]=\frac{\sqrt{-g}}{8\pi G}(d^{D-2}x)_{ab}\left(\chi^a\nabla_ch^{bc}-\chi^a\nabla^bh+\chi_c\nabla^bh^{ac}+\frac{1}{2}h\nabla^b\chi^a-h^{cb}\nabla_c\chi^a\right),
\end{equation}
where $h=g^{ab}h_{ab}$ and $(d^{D-2}x)_{ab}=\frac{1}{2(D-2)!}\epsilon_{a bc_1\ldots c_{D-2}}dx^{c_1}\wedge\ldots\wedge dx^{c_{D-2}}$. The $\ndelta$ is a notation that emphasizes the fact that the charges \eqref{deltaQ} are a priori non-integrable (namely not $\delta$-exact). In the integrable case, $Q_{\chi}$ represents the generator of the associated infinitesimal transformation $\chi$. Computing $\ndelta Q[g,h]$ for the metric written in the horizon gauge \eqref{metric}, the associated preserving vector fields \eqref{Killing} and integrated on a spatial section of $\mathcal{H}$, one obtains \cite{Donnay:2016ejv}:
\begin{equation}
\badat{2}\label{Qphase}
\ndelta Q_{(f,Y^A)}[g, \delta g]=\frac{1}{16\pi G}\oint_{S^{D-2}}d^{D-2}x\,&\biggl(2f\kappa\delta\sqrt{\Omega}+2\partial_vf\delta\sqrt{\Omega}-2f\sqrt{\Omega}\delta\Theta+\frac{1}{2}f\sqrt{\Omega}\partial_v\Omega_{AB}\delta\Omega^{AB}\\
&-Y^A\delta(\theta_A\sqrt{\Omega})\vphantom{12}\biggr).
\eadat
\end{equation}
We can see that these charges are not integrable in full generality, due to the presence of the three following terms: $2f\sqrt{\Omega}\delta\Theta$, $2f\kappa\delta\sqrt{\Omega}$ and $\frac{1}{2}f\sqrt{\Omega}\partial_v\Omega_{AB}\delta\Omega^{AB}$. The authors of \cite{Donnay:2016ejv} circumvent this issue by restricting the phase space to the configurations where $\kappa$ is a constant. They also use the fact that they work in four dimensions to choose a spatial metric related to the usual metric on the 2-sphere by a Weyl transformation. We would like instead for the moment to keep all possible dependencies of the fields.

When surface charges are non-integrable, there is still a way to obtain a representation of the asymptotic Killing algebra through the definition of a modified bracket \cite{Barnich:2011mi}. To do so, we split $\ndelta Q_{\chi}$ into an integrable part $Q^{\mathrm{int}}_{\chi}$ and a non-integrable part $\Xi_{\chi}$:
\begin{equation}
\ndelta Q_{\chi}[g,\delta g]=\delta( Q^{\mathrm{int}}_{\chi}[g])+\Xi_{\chi}[g,\delta g],
\end{equation}
where 
\begin{equation}\label{int}
Q^{\mathrm{int}}_{\chi}[g]=\frac{1}{16\pi G}\oint_{S^{D-2}}d^{D-2}x\sqrt{\Omega}\,\left(2f\kappa+2\partial_vf-\frac{2}{D-2}f\Theta-Y^A\theta_A\right),
\end{equation}
and
\begin{equation}\label{nint}
\Xi_{\chi}[g,\delta g]=-\frac{1}{8\pi G}\oint_{S^{D-2}}d^{D-2}x\sqrt{\Omega}\,f\left(\delta\kappa+\frac{D-3}{D-2}\delta \Theta-\frac{1}{2}\sigma_{AB}\delta \Omega^{AB}\right).
\end{equation}
From this splitting\footnote{This splitting coincides with expression obtained in \cite{Hopfmuller:2018fni} in the Hamiltonian framework, while another splitting was considered in \cite{Donnay:2016ejv} for the case $\kappa=cst$.} we can see directly why, for three-dimensional bulk spacetimes, the condition $\delta \kappa=0$ considered in \cite{Donnay:2016ejv} was sufficient to insure integrability of the charges (the shear vanishes by definition and the factor $(D-3)$ cancels the contribution of the expansion in \eqref{nint}). We now define the following modified Dirac bracket 
\begin{equation}\label{rep}
\{Q^{\mathrm{int}}_{\chi}[g],Q^{\mathrm{int}}_{\eta}[g]\}^{*}\equiv \delta_{\eta}Q^{\mathrm{int}}_{\chi}[g]+\Xi_{\eta}[g,\mathcal{L}_{\chi} g].
\end{equation}
It was first introduced in \cite{Barnich:2011mi} for the study of the BMS charges in four dimensions, which are also generically non-integrable. They also noticed that the splitting is not unique in the sense that for some $N_{\chi}[g]$ we can always choose 
\begin{equation}
\tilde{Q}^{\mathrm{int}}_{\chi}=Q^{\mathrm{int}}_{\chi}-N_{\chi}\quad\text{with}\quad \tilde{\Theta}_{\chi}+\delta N_{\chi}.
\label{FreedomSplitting}
\end{equation}
However, we will see that the separation \eqref{int}, \eqref{nint} we have chosen happens to be relevant in the Carrollian anaysis that we perform in Sec. \ref{III}. This modified bracket defines a representation of the asymptotic Killing algebra: indeed, letting $(f_1,Y_1^A)$ and $(f_2, Y_2^A)$ to be two asymptotic Killing fields, one can show that
\begin{equation}
\{Q^{\mathrm{int}}_{(f_1,Y_1^A)},Q^{\mathrm{int}}_{(f_2,Y_2^A)}\}^{*}=Q^{\mathrm{int}}_{(f_{12}, Y_{12}^A)},
\label{algebra}
\end{equation}
where $f_{12}=f_1\partial_vf_2-f_2\partial_vf_1+Y_1^A\partial_Af_2-Y_2^A\partial_Af_1$ and $Y_{12}^A=Y_1^B\partial_BY_2^A-Y_2^B\partial_BY_1^A$. We notice that this algebra does not involve any central extension. A direct consequence of \eqref{algebra} is that the non-integrable part of the charges plays the role of a source for the non-conservation of $Q^{\mathrm{int}}$. Indeed choosing $(f_2,Y_2^A)$ to be $(1,0)$ we obtain
\begin{equation}
\delta_{(1,0)}Q^{\mathrm{int}}_{\chi}[g]+Q^{\mathrm{int}}_{(\partial_vf,0)}[g]=-\Xi_{(1,0)}[g,\mathcal{L}_{\chi}g],
\end{equation}
moreover $\delta_{(1,0)}$ acts like a time derivative on the fields \eqref{deltachi}, so we finally obtain
\begin{equation}
\frac{d}{dv} Q^{\mathrm{int}}_{\chi}[g]=-\Xi_{(1,0)}[g,\mathcal{L}_{\chi}g].
\end{equation}

\section{Near-horizon or ultra-relativistic limit}\label{III}

One of the particularity of null hypersurfaces is that they are equipped with a degenerate induced metric $\Omega$ in the sense that there exists a vector field $\vec{u}$ that belongs to its kernel:
\begin{equation}
\Omega(.,\vec{u})=0.
\end{equation}
In the case of the horizon described above, $\Omega=\Omega_{AB}(v,\mathbf{x})dx^Adx^B$ and $\vec{u}=f(v,\mathbf{x})\partial_v$, for any function $f$ on $\mathcal{H}$. It was understood,  for example in \cite{Hartong:2015xda,Duval:2014lpa,Penna:2018gfx}, that this defines a \emph{Carrollian geometry}, the natural non-Riemannian geometry that ultra-relativistic theories couple to. This means that any null hypersurface can be thought of as an ultra-relativistic spacetime. In particular, for the near-horizon geometry presented above, we are going to show that the limit $\rho\rightarrow 0$, can be understood as an ultra-relativistic limit where $\sqrt{\rho}$ plays the role of a virtual velocity of light $c$. Notice that this parameter should not be confused with the physical velocity of light of the bulk spacetime that is set to 1 in \eqref{metric}.

This feature has strong consequences on the dynamics of the horizon, \emph{i.e.} the null Raychaudhuri and Damour equations: indeed, we will show that they match ultra-relativistic conservation laws written in terms of the Carrollian geometry and the \emph{Carrollian momenta}, sort of ultra-relativistic equivalent of the energy--momentum tensor.

Finally, we will study the symmetries and charges associated with the horizon that we interpret as \emph{Carrollian Killing}, defined as the vector fields on $\mathcal{H}$ that preserve the Carrollian geometry. In some instances, the symmetry algebra will be shown to have a BMS-like structure in the sense that it includes superrotations and supertranslations on the horizon \cite{Donnay:2015abr,Donnay:2016ejv,Chandrasekaran:2018aop}.

\subsection{Carrollian geometry:  Through the Looking-Glass}\label{III.1}

Carrollian geometry emerges from an ultra-relativistic ($c \to 0$) limit of the relativistic metric and was shown to have a rich mathematical structure and interesting dynamics \cite{Leblond,Duval:2014uva, Duval:2014uoa,Duval:2014lpa,Ciambelli:2018xat,Ciambelli:2018wre,Ciambelli:2018ojf}. It was shown in \cite{Ciambelli:2018ojf, Ciambelli:2018xat} that the $c\rightarrow 0$ limit of relativistic general-covariant theories is covariant under a subset of the diffeomorphisms dubbed \emph{Carrollian diffeomorphisms}
\be 
\label{Cdiff}
v'=v'(v,\bold x) \virg \bold x'=\bold x'(\bold x),
\en
whose infinitesimal version is given by the vector fields
\begin{equation}
\xi=f(v,\mathbf{x})\partial_v+Y^A(\mathbf{x})\partial_A,
\end{equation}
for any $f$ and $Y^A$. This suggests that space and time decouple  and an adequate parametrization to study the ultra-relativistic limit is the so-called Randers--Papapetrou parametrization, where the metric is decomposed as\footnote{Any spacetime metric can be parametrized in that way.}
\be
a=\left( {\begin{array}{cc}
   -c^2\alpha^2 & c^2 \alpha b_A \\
   c^2 \alpha b_B  & \Omega_{AB}-c^2 b_A b_B \\
  \end{array} } \right)_{\{dv,dx^A\}} \underset{c\rightarrow 0}{\longrightarrow}\quad\Omega_{AB}dx^Adx^B.
\label{RP}
\en
After the limit is performed, one thus trade the metric $a$ for $\alpha(v,\mathbf{x})$ the time lapse, $b_A(v,\mathbf{x})$ the temporal connection, and $\Omega_{AB}(v,\mathbf{x})$ the spatial metric. These functions define the Carrollian geometry and one can check that they transform covariantly under Carrollian diffeomorphisms (see Sec. 2 of \cite{Ciambelli:2018ojf} for a complete presentation). Out of the Carrollian geometry, one can build the following first-derivative quantities:
\be
\badat{2}\label{geom}
&\varphi_A = \alpha^{-1}(\p_v b_A +\p_A \alpha),\\
&\beta = \alpha^{-1}\p_v \ln \sqrt \Omega,\\
&\xi_{AB}= \alpha^{-1}\left(\frac{1}{2}\p_v \Omega_{AB} -\frac{\Omega_{AB}}{D-2} \p_v \ln \sqrt \Omega\right),\\
&\omega_{AB}=\p_{[A}b_{B]}+\alpha^{-1}(b_{[A}\p_{B]}\alpha+b_{[A}\p_v b_{B]});
\eadat
\en
they are respectively, the Carrollian acceleration, expansion, shear and vorticity. They also transform covariantly under Carrollian diffeomorphisms, and will play an important role in the Carrollian conservation laws we will discuss in the next section.

Let us come back to the black hole near-horizon metric \eqref{metric}. On each constant $\rho$ hypersurface, called $\Sigma_{\rho}$ in Fig. \ref{ourfigure}, it induces a Lorentzian signature metric that becomes degenerate when taking the near-horizon limit:
\be\label{h}
a=ds^2_{\rho=cst}=\begin{pmatrix}
-2\rho\kappa & \rho \theta_A\\
\rho \theta_B & \Omega_{AB}+\rho \lambda_{AB}
\end{pmatrix}_{\{dv,dx^A\}} \underset{\rho\rightarrow 0}{\longrightarrow}\quad\Omega_{AB}dx^Adx^B.
\en
If we now compare this induced metric with the Randers--Papapetrou one, we are tempted to make the following identifications:\footnote{One notices that, following this identification, we should also have $\lambda_{AB}=-b_Ab_B$, which becomes $\lambda_{AB}=-\frac{\theta_A\theta_B}{2\kappa}$. This would then impose a constraint on the near-horizon geometry, but we will actually not have to do that as $\lambda_{AB}$ will always appear at subleading order in the equations we are going to consider.}
\begin{equation}
c^2=\rho,\quad \alpha=\sqrt{2\kappa},\quad\text{and}\quad b_A=\frac{\theta_A}{\sqrt{2\kappa}}.
\label{identification}
\end{equation}
We thus identify the radial coordinate with the square of a virtual speed of light for the Lorentzian spacetime $\Sigma_{\rho}$. As the horizon is located at $\rho=0$, it is an ultra-relativistic spacetime endowed with a Carrollian geometry given in terms of the surface gravity, the twist and the induced spatial metric $\Omega_{AB}$. 
After this identification, we can re-express the first-derivative Carrollian tensors \eqref{geom} in terms of the extrinsic geometry of the horizon \eqref{extrinsic}:
\begin{equation}
\badat{3}\label{geom2}
&\varphi_A =\frac{1}{2\kappa}\left(\partial_A\kappa+\partial_v\theta_A-\frac{\theta_A}{2\kappa}\partial_v\kappa\right),\\
&\beta=\frac{\Theta}{\sqrt{2\kappa}},\\
&\xi_{AB}=\frac{1}{\sqrt{2\kappa}}\sigma_{AB},\\
&\omega_{AB}=\frac{1}{2}\left(\frac{\p_A \theta_B}{\sqrt{2\kappa}}+\frac{2\theta_A \p_B \kappa+\theta_A \p_v \theta_B}{(2\kappa)^{3/2}}\right)-(A\leftrightarrow B).\\
\eadat
\end{equation}
We notice that the Carrollian expansion and the Carrollian shear are proportional respectively to the expansion and the shear of the horizon defined extrinsically in Sec. \ref{section2.1}.

\subsection{Horizon dynamics as ultra-relativistic conservation laws}\label{III.2}

We now turn our attention to the gravitational dynamics of the horizon. Consider again the hypersurface $\Sigma_{\rho}$ near $\rho=0$. Its unit normal is given by
\begin{equation}
n=\frac{d\rho}{\sqrt{2\kappa\rho}},
\label{normal}
\end{equation}
and allows us to define the extrinsic curvature  and the momentum conjugate to the induced metric: 
\be
T_{ab}=\frac{1}{8\pi G}(K a_{ab}-K_{ab}),
\label{BYtensor}
\en
where $K^a_b=a^c_b \mathcal{D}_c n^a$ is the extrinsic curvature of $\Sigma_{\rho}$, $K=K^a_a$ its trace and $a_{ab}=g_{ab}-n_a n_b$ is the projector on the hypersurface perpendicular to $n$.\footnote{The projector $a_{ab}$ coincides with \eqref{h} when one consider its $\{v,A\}$ components only.} This hypersurface is sometimes referred to as the stretched horizon or membrane, while $T_{ab}$ is called the ``membrane energy--momentum tensor'' \cite{Thorne:1986iy,Price:1986yy,Penna:2015gza}.\footnote{In those papers, the approach is to study this membrane energy--momentum tensor for a small $\rho$ and use it to define the fluid quantities like the energy density, the pressure, etc. The problem is that those quantities diverge when $\rho$ is sent to zero. Their solution is to rescale them by hand to obtain finite quantities. We propose another approach and define the Carrollian momenta that are finite on the horizon and well suited for the ultra-relativistic interpretation.} Einstein equations ensure that it is conserved:
\begin{equation}
\bar{\nabla}_jT^{ji}=0,
\label{Cons}
\end{equation}
where the index $i$ refers to $\{v,\mathbf{x}\}$, and $\bar{\nabla}_i$ is the Levi-Civita connection associated with the induced metric \eqref{h}. The membrane is then interpreted as a fluid whose equations of motion are given by this conservation law. One notices that \eqref{Cons} describes the dynamics of a \emph{relativistic} fluid that lies in the $(D-1)$-dimensional spacetime given by the constant $\rho$ hypersurface and equipped with the metric $a$. We are going to show that, to obtain the null Raychaudhuri \eqref{Ray} and Damour equations \eqref{Damour}, one has to take the near-horizon limit of this conservation law which, at the level of the fluid, is interpreted as an ultra-relativistic limit through the identification $\rho=c^2$.

Using \eqref{metric}, we compute the membrane energy--momentum tensor near the horizon,
\begin{equation}
\badat{3}\label{Tab}
& 8\pi GT^{vv}=\frac{\Theta}{2\sqrt{2}(\rho \kappa)^{\frac{3}{2}}}+\mathcal{O}(1/\sqrt \rho),\\
&8\pi GT^{vA}=-\frac{1}{2\sqrt{2\rho}\kappa^{3/2}}\left(\p^A \kappa+\theta^A( \kappa+\Theta)\right)+\mathcal{O}(\sqrt \rho),\\
&8\pi GT^{AB}=-\frac{1}{ \sqrt{2 \rho \kappa}}\left(\Omega^{AB}(\kappa+\Theta-\frac{\p_v \kappa}{2\kappa})+\frac{1}{2}\p_v \Omega^{AB}\right)+\mathcal{O}(\sqrt \rho).
\eadat
\end{equation}
We now decompose $T^{ij}$ into the \emph{Carrollian momenta}, which are defined such that they are independent of the speed of light and covariant under Carrollian diffeomorphisms \cite{Ciambelli:2018ojf},
\begin{equation}
\badat{3}\label{dec}
&8\pi G T^{vv}=c^{-3} \alpha^{-2}\mathcal{E}+\mathcal{O}(c^{-1}) ,\\
& 8\pi G T^{vA}=c^{-1}\alpha^{-1}(\pi^{A}-2b_{B}\mathcal{A}^{AB})+\mathcal{O}(c),\\
&8\pi G T^{AB}=-2 c^{-1}\mathcal{A}^{AB}+\mathcal{O}(c),
\eadat
\end{equation}
with $\rd E$ a scalar, $\pi^A$ a spatial vector and $\mathcal{A}^{AB}$ a spatial symmetric 2-tensor. They are the ultra-relativistic equivalent of an energy--momentum tensor. They can be thought of respectively as the energy density, the heat current and the total stress tensor. The latter can be decomposed into its trace and traceless part
\begin{equation}
\mathcal{A}^{AB}=-\frac{1}{2}\left(\mathcal{P}\Omega^{AB}-\Xi^{AB}\right),
\end{equation}
which are interpreted respectively as the pressure and the dissipative tensor.

Comparing \eqref{Tab} with \eqref{dec}, we read the following Carrollian momenta: 
\begin{equation}\label{Cmom}
\begin{split}
&\mathcal{E}=\frac{1}{\sqrt{2\kappa}}\Theta,\\
&\mathcal{P}=-\frac{1}{\sqrt{2\kappa}}\left(\kappa+\frac{D-3}{D-2}\Theta-\frac{\partial_v\kappa}{2\kappa}\right),\\
&\Xi_{AB}=-\frac{1}{\sqrt{2\kappa}}\sigma_{AB},\\
&\pi_A=-\frac{1}{2}\left(\frac{\partial_A\kappa}{\kappa}+\frac{\theta^B}{2\kappa}\partial_v\Omega_{BA}+\frac{\theta_A}{2\kappa^2}\partial_v\kappa\right).
\end{split}
\end{equation}
We have obtained that the energy density is proportional to the expansion of the horizon. The pressure is related to the combination
\begin{equation}
\mu=\kappa+\frac{D-3}{D-2}\Theta,
\label{mu}
\end{equation}
which is referred to in \cite{Hopfmuller:2018fni} as the ``gravitational pressure'' and receives corrections from the time evolution of the surface gravity. The dissipative tensor is proportional to the shear of the horizon \eqref{shear}. The heat current $\pi_A$ is harder to interpret but we notice that it receives a contribution from the gradient of $\kappa$, which can be thought of as a local temperature on the black hole horizon (see the discussion at the end of \cite{Kirklin:2018wvq}). 

These Carrollian momenta satisfy conservation equations that are given by the ultra-relativistic (\emph{i.e.} near-horizon) limit of the energy--momentum conservation  \eqref{Cons}.\footnote{This limit was considered for the first time for a relativistic fluid in \cite{Ciambelli:2018xat}.} Using the decompositions for the metric \eqref{RP} and the energy--momentum tensor \eqref{dec}, we obtain:
\begin{equation}
\begin{split}\label{Cce}
&\left(\alpha^{-1}\partial_v+\beta\right)\mathcal{E}-\rd A^{AB}\alpha^{-1}\partial_v \Omega_{AB}=0,\\
&2\left(\hat{\nabla}_A+\varphi_A\right)\mathcal{A}^A_B-\mathcal{E}\varphi_B-\left(\alpha^{-1}\partial_v+\beta\right)\pi_B=0.
\end{split}
\end{equation}
These equations\footnote{These conservation equations were also shown to reproduce the constraint equations on the null infinity for asymptotically flat spacetimes in the Bondi gauge, see \cite{Ciambelli:2018ojf}.} are covariant w.r.t. Carrollian diffeomorphisms, in the sense that the first one transforms like a scalar and the second one like a spatial vector and they are independant of $c$ (or $\rho$, the radial coordinate). We have introduced a new object $\hat{\nabla}_A$, which is a Carroll-covariant derivative:
\begin{equation}
\hat{\nabla}_{A}v^B=\hat{\partial}_Av^B+\hat{\gamma}^A_{BC}v^C,
\end{equation}
where
\begin{equation}
\hat{\partial}_A=\partial_A+\frac{b_A}{\alpha}\partial_v\quad\text{and}\quad \hat{\gamma}^A_{BC}=\frac{1}{2}\Omega^{AD}\left(\hat{\partial}_B\Omega_{DC}+\hat{\partial}_C\Omega_{DB}+\hat{\partial}_D\Omega_{BC}\right).
\end{equation}
If $v^A$ transforms like a spatial vector, \emph{i.e.} $v'^A=\frac{\partial x'^A}{\partial x^B}v^B$ under a Carrollian diffeomorphism \eqref{Cdiff}, then $\hat{\nabla}_Av^B$ will transform like a spatial $2$-tensor. One can check that this would not be the case for the usual Levi-Civita connection associated with $\Omega_{AB}$. The first equation of \eqref{Cce} can be interpreted as a conservation of energy on a curved background, but an exotic one: indeed, one would expect the gradient of the heat current to appear while here it is absent even when the heat current is non zero. This feature is a signature of the ultra-relativistic limit \cite{Ciambelli:2018xat}.

The main result of this section is that, considering the Carrollian geometry \eqref{identification} and the Carrollian momenta \eqref{Cmom} and after a lenghty computation, one can show that the scalar equation is exactly the null Raychaudhuri equation \eqref{Ray} while the spatial one gives the Damour equation \eqref{Damour}. This confirms that the dynamics of a black hole is mapped to ultra-relativistic conservation laws when the near-horizon radial coordinate is identified with a virtual speed of light.

\subsection{Conserved charges on the horizon}

Using the results of the previous section we would like now to build conserved charges associated with the horizon. The idea is to use the techniques we know from relativistic physics to build charges on a constant $\rho$ hypersurface and then send the radial coordinate to zero to obtain conserved charges on the horizon.  The latter will be conserved on shell and associated to the symmetries of the induced Carrollian geometry on the horizon. At the end of this section, we discuss their relationship with the one obtained through the covariant phase space formalism in Sec. \ref{sec:bulkcharges}. 

\subsubsection*{Charges associated to Carrollian Killing fields on the horizon}

Consider again the energy--momentum tensor of the membrane \eqref{BYtensor}: vacuum Einstein equations imply that it is conserved:
\begin{equation}
\bar{\nabla}_j T^{ji}=0.
\end{equation} 
It is thus possible to build a conserved current associated with any vector field of $\Sigma_{\rho}$ that satisfies the Killing equation for the induced metric $a_{ij}$:
\begin{equation}
\bar{\nabla}_i\xi_j+\bar{\nabla}_j\xi_i=0,
\label{KillingCarroll}
\end{equation}
where we recall that $\bar{\nabla}_i$ is the Levi-Civita associated with $a$. This current is given by $J^i=\xi_jT^{ji}$; it is conserved
\begin{equation}
\bar{\nabla}_i J^i=0,
\end{equation}
and allows to build, for any small $\rho$, a conserved charge w.r.t. the $v$ coordinate:
\begin{equation}
\mathcal{Q}_{\xi}^{\rho}=\oint_{S_{v,\rho}} d^{D-2}x \sqrt{q}\ \ell_i J^i,
\label{charge-rho}
\end{equation}
where
\begin{equation}
q_{AB}=\Omega_{AB}+\rho\lambda_{AB}+\mathcal{O}(\rho^2)\quad\text{and}\quad \ell=\sqrt{2\kappa\rho}\ dv+\mathcal{O}(\rho^{\frac{3}{2}}),
\end{equation} 
are respectively the induced metric on a spatial section of the constant $\rho$ hypersurface, \emph{i.e.} $S_{v,\rho}$, and the unit timelike normal to the spatial section in the constant $\rho$ hypersurface, see Fig. \ref{ourfigure}.

We are now ready to perform the near-horizon limit of this construction. We consider first the Killing equation for the vector $\xi$ that we decompose as $\xi=f(v,\mathbf{x})\partial_v+Y^A(v,\mathbf{x})\partial_A$. The zero-$\rho$ limit of \eqref{KillingCarroll} becomes
\begin{equation}
\begin{split}
&\partial_vY^A=0,\\
&f\partial_v\kappa+Y^A\partial_A\kappa+2\kappa\partial_vf=0,\\
&f\partial_v  \Omega_{AB}+\nabla_AY_B+\nabla_BY_A=0.
\end{split}
\label{CK}
\end{equation}
The first thing to notice is that the near-horizon limit of the Killing equation imposes the vector field $\xi$ to be Carrollian! Moreover, these three equations have an interesting geometrical interpretation: indeed, consider the degenerate metric induced on the horizon $\Omega=\Omega_{AB}(v,\mathbf{x})dx^Adx^B$ and the vector field $\vec{v}=\alpha^{-1}\partial_v$ (where $\alpha$ is given by the identification \eqref{identification}), they are equivalent to asking
\begin{equation}
\mathcal{L}_{\xi}\vec{v}=0\quad\text{and}\quad \mathcal{L}_{\xi}\Omega=0.
\label{CK2}
\end{equation}
Following \cite{Duval:2014lpa}, the triple ($\mathcal{H}$, $\Omega$, $\vec{v}$) defines a non-Riemannian geometry called \emph{weak Carroll manifold}.\footnote{The Carrollian geometry also involves the temporal connection $b_A$ but is does not appear in the definition of Carrollian Killings.} The latter is the natural structure that appears when one wants to study ultra-relativistic symmetries. Things appear to be consistent: we have considered the symmetries of the relativistic metric $a$, \emph{i.e.} its Killing vector fields, then we have taken the near-horizon limit, interpreted as an ultra-relativistic limit for $\rho=c^2$, and we obtain the symmetries of the corresponding Carrollian geometry. These symmetries given by Eq. \eqref{CK2} will be called \emph{Carrollian Killing} symmetries.

We can also perform the near-horizon limit of the charge \eqref{charge-rho} using the value of the membrane energy--momentum tensor derived in Sec. \ref{III.2}; we obtain
\begin{equation}
\mathcal{Q}^{\rho}_{\xi}\underset{\rho\rightarrow 0}{\longrightarrow}\mathcal{C}_{\xi}=\frac{1}{16\pi G}\oint_{S^{D-2}}\text{d}^{D-2}x\sqrt{\Omega}\left(-2f\Theta-Y^A\left(\theta_A+\frac{\partial_A\kappa}{\kappa}\right)\right).
\end{equation}
This charge is conserved provided that the null Raychaudhuri and the Damour equations are satisfied and the couple $(f,Y^A)$ satisfies the Carrollian Killing equations \eqref{CK}. Taking the trace of the last equation of \eqref{CK} we obtain $f\Theta=-\nabla_AY^A$, therefore the integration on the sphere of this term vanishes. The charge becomes
\begin{equation}
\mathcal{C}_{\xi}=\frac{-1}{16\pi G}\oint_{S^{D-2}}\text{d}^{D-2}x\sqrt{\Omega}Y^A\left(\theta_A+\frac{\partial_A\kappa}{\kappa}\right).
\end{equation}
This is a sort of generalization of the angular momentum to the case of non-stationary black holes. We would like indeed to stress that in this formula, $\Omega_{AB}$, $\kappa$ and $\theta_A$ depend generically on both $v$ and $x^A$, so the conservation of this charge is really non-trivial. Therefore, to any isometry of the induced Carrollian geometry on the horizon, we have associated a charge that is conserved on-shell.

 When we consider the case $\kappa=cst$ and $\Omega_{AB}=\bar{\Omega}_{AB}(\mathbf{x})$, the solutions to the Carrollian Killing equations are a supertranslation $f=\mathcal{T}(\mathbf{x})$ together with a real Killing of the metric $\bar{\Omega}_{AB}$ and if one considers the near-horizon geometry of a Kerr black hole and the spatial Killing $Y=\partial_{\varphi}$, this charge reproduces the constant angular momentum $J$ \cite{Donnay:2016ejv}.

 \subsubsection*{The conformal case}
 
The same analysis can be carried out for a conformal Killing on the constant $\rho$ hypersurface $\Sigma_{\rho}$, \emph{i.e.} a vector $\xi$ that satisfies
\begin{equation}
\bar{\nabla}_i\xi_j+\bar{\nabla}_j\xi_i=2\lambda a_{ij},
\label{ConfKilling}
\end{equation} 
where $\lambda(v,\mathbf{x})$ is any function. We can build the same current by projecting $\xi$ on the energy--momentum tensor. However, if $\lambda\neq 0$, the associated charge will be conserved on-shell only if $T^{ij}$ satisfies the tracelessness condition 
\begin{equation}
T^i_i=0.
\label{Traceless}
\end{equation}
The near-horizon limit of the conformal Killing equation is
\begin{equation}
\begin{split}
&\partial_vY^A=0,\\
&f\partial_v\kappa+Y^A\partial_A\kappa+2\kappa\partial_vf=2\kappa\lambda,\\
&f\partial_v  \Omega_{AB}+\nabla_AY_B+\nabla_BY_A=2\lambda\Omega_{AB}.
\end{split}
\label{CCK}
\end{equation}
Again, it admits a nice interpretation as the conformal isometries of the weak Carroll manifold induced on the horizon. Indeed, \eqref{CCK} is equivalent to 
\begin{equation}
\mathcal{L}_{\xi}\vec{v}=-\lambda\vec{v}\quad\text{and}\quad\mathcal{L}_{\xi}\Omega=2\lambda\Omega,
\end{equation} 
and, according to \cite{Duval:2014lpa}, this is the definition of the level-2 conformal isometries of $(\mathcal{H}, g, \vec{v})$; we will call them \emph{conformal Carrollian Killing} vectors. To any conformal Carrollian Killing $\xi$ we can associate the following charge:
\begin{equation}
\mathcal{C}_{\xi}=\frac{1}{16\pi G}\oint_{S^{D-2}}\text{d}^{D-2}x\sqrt{\Omega}\left(-2f\Theta-Y^A\left(\theta_A+\frac{\partial_A\kappa}{\kappa}\right)\right),
\end{equation}  
which is the same as in the previous section, obtained through the near-horizon limit of $\mathcal{Q}_{\xi}$. The only difference is that, if $\lambda\neq 0$, this charge will not be generically conserved on-shell. It is generically conserved only if the near-horizon limit of the tracelessness condition \eqref{Traceless} is satisfied, \emph{i.e.}
\begin{equation}
\mathcal{S}\equiv \Theta+\kappa-\frac{\partial_v\kappa}{2\kappa}=0,
\end{equation}
where the function $\mathcal{S}$ has been defined through
\begin{equation}
T^i_i\underset{\rho\rightarrow 0}{\longrightarrow}\frac{-1}{8\pi G \sqrt{2\kappa}\sqrt{\rho}}\mathcal{S}.
\end{equation}
Asking $\mathcal{S}$ to be zero is a non-trivial additional constraint on the surface gravity and the expansion, that we will call the \emph{conformal state equation}. Indeed, if we reintroduce the Carrollian momenta \eqref{Cmom} we obtain that
\begin{equation}
\mathcal{S}=0\quad\Leftrightarrow\quad\mathcal{E}=(D-2)\mathcal{P}.
\label{Cstate}
\end{equation}
We recognize the usual state equation satisfied by the energy and the pressure of a conformal fluid (see \cite{Rangamani:2009xk} or \cite{Ciambelli:2018xat}). 

We consider now the case $\kappa=cst$ and $\Omega_{AB}=\bar{\Omega}_{AB}(\mathbf{x})$, the corresponding Carrollian Killings are given by
\begin{equation}
\xi=\left(\frac{v}{D-2}\nabla_AY^A+\mathcal{T}(\mathbf{x})\right)+Y^A(\mathbf{x})\partial_A,
\label{bms}
\end{equation}
where $\mathcal{T}$ is a supertranslation and $Y^A$ is a conformal Killing of $\bar{\Omega}_{AB}$. When the spatial metric is chosen to be the round metric on $S^{D-2}$ we obtain the $bms_D$ algebra. The conformal state equation becomes $\kappa=0$. This constraint is obviously very restricting but actually, in this particular case, we will not have to impose it to obtain conserved charges. Indeed the charge $\mathcal{C}_{\xi}$ becomes
\begin{equation}
\mathcal{C}_{\xi}=\frac{-1}{16\pi G}\oint_{S^{D-2}}\text{d}^{D-2}x\sqrt{\bar{\Omega}}Y^A\theta_A,
\end{equation}
and the Damour equation becomes
\begin{equation}
\partial_v\theta_A=0.
\end{equation}
So, for any value of $\kappa$, this charge associated to a conformal Carrollian Killing of the type \eqref{bms} is manifestly conserved on-shell, but insensitive to the supertranlsation $\mathcal{T}$. 

\subsubsection*{Relationship with the bulk analysis}

Finally, in both the non-conformal and conformal case, we can relate $\mathcal{C}_{\xi}$ to the integrable part of the charges obtained through the covariant phase space formalism in Sec. \ref{sec:bulkcharges}. Indeed, consider an asymptotic Killing $(f, Y^A)$ \eqref{Killing}; as already stated in Sec. \ref{sec:bulkcharges}, its projection on the horizon is a generic Carrollian vector field. We can further ask the latter to be a (conformal-)Carrollian Killing, thus considering the subset of asymptotic Killings whose projection on the horizon provides an isometry of the induced Carrollian geometry. If we do so, one can show that
\begin{equation}
\mathcal{C}_{(f,Y^A)}=Q^{\mathrm{int}}_{(f,Y^A)}-\frac{1}{8\pi G}\oint_{S^{D-2}}d^{D-2}x\sqrt{\Omega}f\mathcal{S},
\end{equation}
where we notice the mysterious appearance of the function $\mathcal{S}$ that defines the conformal state equation \eqref{Cstate}. This equation holds up to boundary terms that are vanishing when integrated on the sphere and if the couple $(f,Y^A)$ satisfies the Carrollian Killing equations \eqref{CK} or its conformal version \eqref{CCK}. This equality is off-shell; if we further impose the equations of motion and perform a time derivative we obtain
\begin{equation}
\frac{d}{dv}Q^{\mathrm{int}}_{(f,Y^A)}=\frac{1}{8\pi G}\oint_{S^{D-2}}d^{D-2}x\sqrt{\Omega}\left[f\partial_v+\partial_vf-\nabla_AY^A\right]\mathcal{S}.
\end{equation}
We conclude that the non-conservation of $Q_{(f,Y^A)}^{\text{int}}$, for (conformal-)Carrollian Killing vectors, will be sourced by the function $\mathcal{S}$. Therefore we have established a connection between the conservation of the charges and the conformality of the Carrollian momenta associated with the horizon. A last remark is that these very compact results are valid for the splitting we have made in Sec. \ref{sec:bulkcharges} between the integrable and non-integrable part of the charge, it would be interesting to determine how they get modified under the change of splitting \eqref{FreedomSplitting}.

\section{Perspectives}\label{IV}

This analysis sets an indubitable connection between Carrollian and near-horizon physics, the main result being that the dynamics of the black hole horizon is given by an ultra-relativistic conservation law. In the membrane paradigm, the ``fluid'' describing the horizon is supposed to satisfy the Damour-Navier Stokes equation, which a priori is a non-relativistic equation but for a Galilean fluid (\emph{i.e.} when the speed of light is infinite). We want to point out that, instead, the fluid behaves more like a Carrollian one.  This observation is emphasized by the fact that the energy conservation satisfied on the horizon seems very different from the one that a usual Galilean fluid would satisfy, as it does not involve the gradient of the heat current (see first equation of \eqref{Cce}), while it is perfectly interpreted in terms of an ultra-relativistic energy conservation. All these remarks lead to the conclusion that the ultra-relativistic approach seems to be more appropriate to the study of horizon dynamics. In \cite{Ciambelli:2018xat}, the authors study the ultra-relativistic limit of a relativistic fluid; it would be interesting to see how this translates in the horizon analysis. One could also study the thermodynamics of such a fluid, especially its entropy current, and see if we can relate it to the black hole entropy.

Another question is the role of the function $\mathcal{S}$ introduced to define the conformal state equation. It would be interesting to understand better its status at the level of the charges. Indeed, the exact same relationship was found in the context of asymptotically flat gravity between the Carrollian charges and the charges obtained through covariant phase space formalism \cite{Ciambelli:2018ojf}. In that case, the function $\mathcal{S}$ (called $\sigma$ there) was representing the flux of gravitational radiation through null infinity and was therefore responsible of the non-conservation of the charges. At the level of the horizon, the function $\mathcal{S}$ could have the same kind of physical interpretation which would be worth clarifying.

Finally, let us mention two other interesting directions. The first one would be to add other fields to source the bulk energy--momentum tensor and see how this analysis get modified, in particular their influence on the charges. The second one is the specific case of extremal black holes. We have not mentioned them in this paper since their study would require strong modifications in our analysis (for instance, Carrollian momenta for $\kappa=0$ would diverge as one can see from \eqref{Cmom}). The study of Carrollian physics for extremal black holes will be the subject of future works.

\section*{Acknowledgments}
We are grateful to L. Ciambelli, G. Giribet, R. Leigh, R. F. Penna, P. M. Petropoulos, and A.-M. Raclariu for useful discussions.
LD acknowledges support from the Black Hole Initiative (BHI) at Harvard University, which is funded by a grant from the John Templeton Foundation. CM thanks the BHI for its hospitality while part of this work was done. This work was also partly funded by the ANR-16-CE31-0004 contract Black-dS-String. 

\bibliographystyle{style}
\bibliography{references}

\end{document}